\documentclass[%
 aip,
 amsmath,amssymb,
 reprint,%
]{revtex4-1}

\usepackage{amsfonts}
\usepackage{amsmath}
\usepackage{amssymb}
\usepackage{graphicx}
\usepackage{epstopdf}
\usepackage{mathrsfs}
\usepackage{color}
\usepackage{bbold}
\usepackage{CJK}
\usepackage{times}
\usepackage[colorlinks, citecolor=red]{hyperref}

\usepackage{graphicx}% Include figure files
\usepackage{dcolumn}% Align table columns on decimal point
\usepackage{bm}% bold math
\usepackage[utf8]{inputenc}
\usepackage[T1]{fontenc}
\usepackage{mathptmx}
\usepackage{color}

\begin{document}

\preprint{AIP/123-QED}

\title[]{Niobium Dayem  nano-bridge Josephson field-effect transistors}
% Force line breaks with \\

\author{G. De Simoni}
\email{giorgio.desimoni@nano.cnr.it}
\affiliation{NEST, Instituto Nanoscienze-CNR and Scuola Normale Superiore, I-56127 Pisa, Italy}
\author{C. Puglia}
%\email{claudio.puglia@df.unipi.it}
\affiliation{Dipartimento di Fisica, Università di Pisa, Largo Bruno Pontecorvo 3, I-56127 Pisa, Italy}
\affiliation{NEST, Instituto Nanoscienze-CNR and Scuola Normale Superiore, I-56127 Pisa, Italy}
\author{F. Giazotto}
\email{francesco.giazotto@sns.it}
\affiliation{NEST, Instituto Nanoscienze-CNR and Scuola Normale Superiore, I-56127 Pisa, Italy}


\begin{abstract}
We report on the first realization of Nb-based \textit{all-metallic} gated Dayem nano-bridge field-effect transistors (Nb-FETs). These Josephson devices operate up to a temperature of $\sim 3$ K, and exhibit full suppression of the  supercurrent thanks to the application of a control gate voltage. The dependence of the kinetic inductance and of the transconductance on gate voltage promises a performance already on par with so far realized metallic Josephson transistors, and let to foresee the implementation of a  superconducting digital logic based on Nb-FETs. We conclude by showing the practical realization of a scheme implementing an all-metallic gate-tunable \emph{half-wave} rectifier to be used either for superconducting electronics or for photon detection applications.
\end{abstract}

\maketitle

\begin{figure}[ht]
\includegraphics[width=0.9\columnwidth]{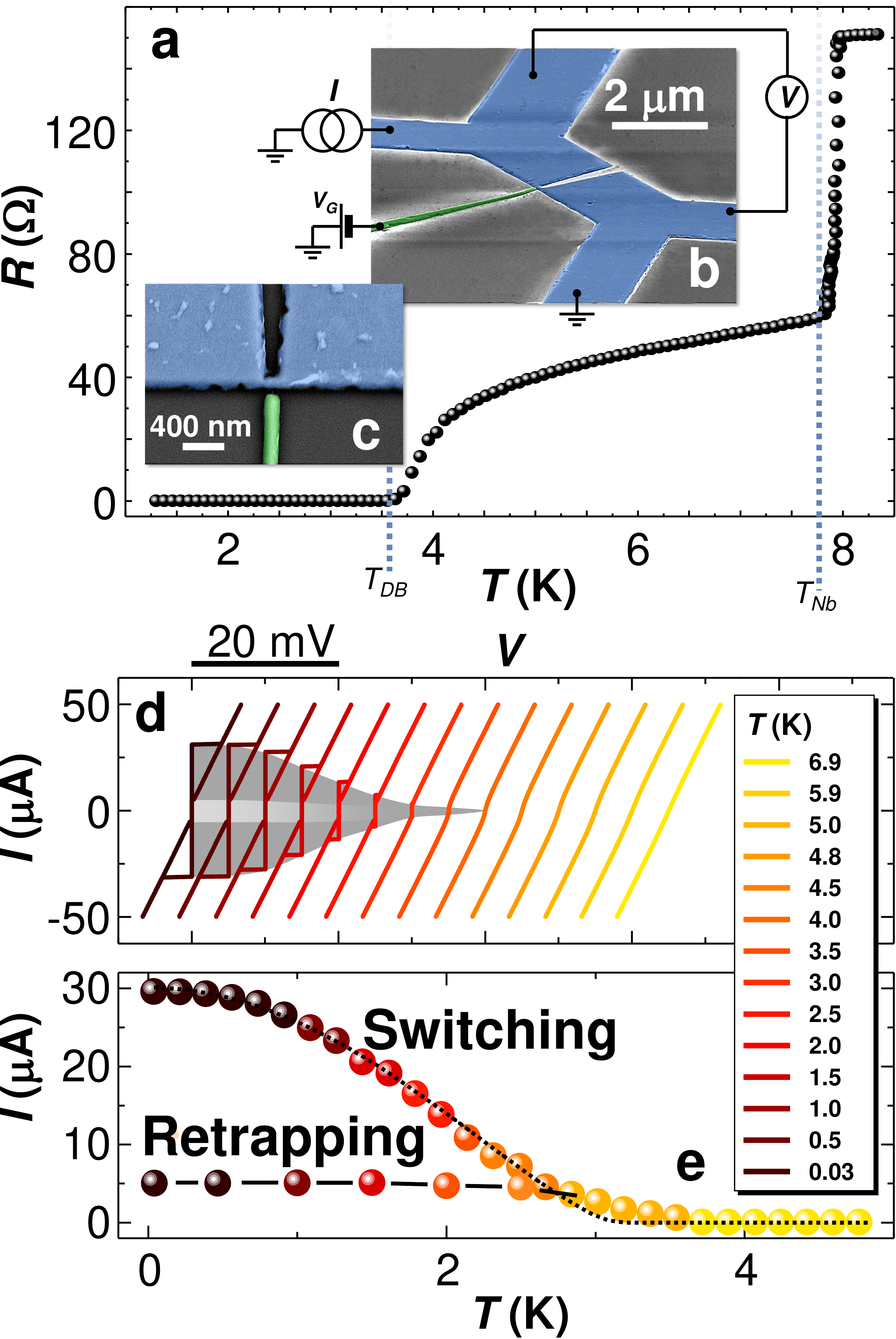}
\caption{\label{fig1} (a) Resistance $R$ \textit{vs.} temperature $T$ characteristic of a representative Nb-FET device. The measurement was performed via standard 4-wire lock-in technique in a filtered dilution refrigerator. 
Two transitions were observed, highlighted by black dashed  lines corresponding to the superconductor-to-normal state transition of the Nb-leads ($T_{Nb}$), and of the DB $(T_{DB}$). 
(b) False color scanning electron micrograph (SEM) of a Nb-FET. The blue area correspond to the Nb leads and the DB. The Nb gate is colored in green. The biasing scheme used for 4-wire dc characterization of our devices is also shown. 
(c) False color SEM blow-up of the DB region. 
(d) Current $I$ \textit{vs.} voltage $V$ characteristics at several bath temperatures $T$ of a representative Nb-FET. Curves are horizontally offset for clarity. 
(e) Switching ($I_S$) and retrapping ($I_R$) currents \emph{vs.} $T$ of the same device of Fig. \ref{fig1}(d). Black dotted line shows the best-fit of the decay of $I_S$ as a function of $T$ with the Bardeen's formula. A guide for the eye is also drown to highlight the decay of $I_R$ with temperature.}
\end{figure}

Superconductor electronics (SCE) deals with electronic circuits based on elements that are
superconducting below their critical temperature ($T_c$), and exhibit unique characteristics and performances which are unrivaled by conventional semiconductor counterparts\cite{Braginski2019,yoshikawa2019superconducting}. SCE relies on the quantum properties of superconductors such as, e.g.,  the Josephson effects \cite{Barone1982,Likharev1986}, the magnetic flux quantization\cite{Likharev1991b,Likharev2012}, and the extremely low-power absorption both in DC and in AC fields up to the superconducting gap
frequency ($f_\Delta=2\Delta / h$ , where $\Delta$ and $h$ are respectively the superconducting gap and the Planck's constant). For practical SCE, the superconducting material of choice is niobium (Nb): a Bardeen-Cooper-Schrieffer (BCS) metal that has the highest $T_c$ ($\sim 9.2$ K) and $f_\Delta$ ($\sim1.4$ THz) among elemental superconductors \cite{Braginski2019}, being therefore suitable for circuit operation at temperatures around $\sim 4$ K. 
Other elemental  superconducting  metals with sizable  $T_c$, such as  vanadium (V)  or  lead (Pb),  are scarcely exploited \cite{Spathis2011,Giazotto2011,paajaste2015pb} for SCE. Furthermore, despite other low-temperature superconductors with lower $T_c$ are widely exploited in radiation detection and for quantum computation architectures, compound superconductors with $T_c$ higher than that of Nb such as, for instance, NbN, carbonitrides and cuprate high-$T_c$ superconductors, have limited SCE applications,  mostly due to complex and expensive film deposition techniques, to the extremely short coherence length and to the anisotropy of their electronic properties\cite{Braginski2019}. 

In this Letter, we report on the first realization of Nb-based \textit{all-metallic} gated Dayem nano-bridge (DB) field-effect transistors (Nb-FETs). Differently from supercurrent \cite{Nishino1989a,Mannhart1993a,Mannhart1993b,Okamoto1992a,Fiory1990a} and Josephson  field-effect  transistors  (SuFETs and JoFET) \cite{Akazaki1995,Doh2005a}, where the critical current of a proximitized semiconductor is controlled via conventional field-effect-driven charge depletion/accumulation, \textit{all-metallic} superconducting transistors (S-FETs) represent a recently-demonstrated class of devices entirely fabricated with BCS metals. In these transistors, the supercurrent flow can be significantly manipulated via electro-static gating \cite{DeSimoni2018,Paolucci2018,Paolucci2019b,Paolucci2019,Paolucci2019a,Paolucci2018} without any variation of the charge density. The most relevant phenomenology observed in S-FETs is the bipolar reduction, down to full-suppression, of  the critical  supercurrent ($I_C$) for both positive and negative gate polarization. Furthermore, the dependence of the phase on an externally applied electrostatic field was recently demonstrated in gated Ti-based superconducting quantum interference devices (SQUIDs) \cite{Paolucci2019a} as well as in the evolution of the switching current probability distributions in Ti DBs \cite{Puglia2019}. 

Although only few theoretical models have attempted to explain the above experiments till now \cite{Mercaldo2019,Bours2020,Virtanen2019a}, the implementation of digital logic gates based on superconducting FETs has already been proposed such as, e. g., AND, COPY, and NOT circuits \cite{Paolucci2019b}. 
All these ports are based on the so-called EF-Tron, i.e., the electric field-effect counterpart of the nano-cryotron  (nTron)\cite{Zhao2017a,Likharev2012a,Buck1956}. The latter is a device where an injection current is used to control the supercurrent flowing in a metallic channel. 
S-FETs based on Al \cite{Bours2020}, Ti \cite{DeSimoni2018}, and V \cite{Paolucci2019b} have already been  demonstrated but no implementation with Nb was reported so far. The results presented here fill this gap, and candidate Nb-FETs as the enabling technology to implement a SCE platform, which is naturally compatible-with and alternative-to both the rapid single flux quantum (RSFQ) \cite{Likharev1991b,Likharev2012}, and the complementary metal-oxide (CMOS) approaches.

\begin{figure}
\includegraphics[width=0.9\columnwidth]{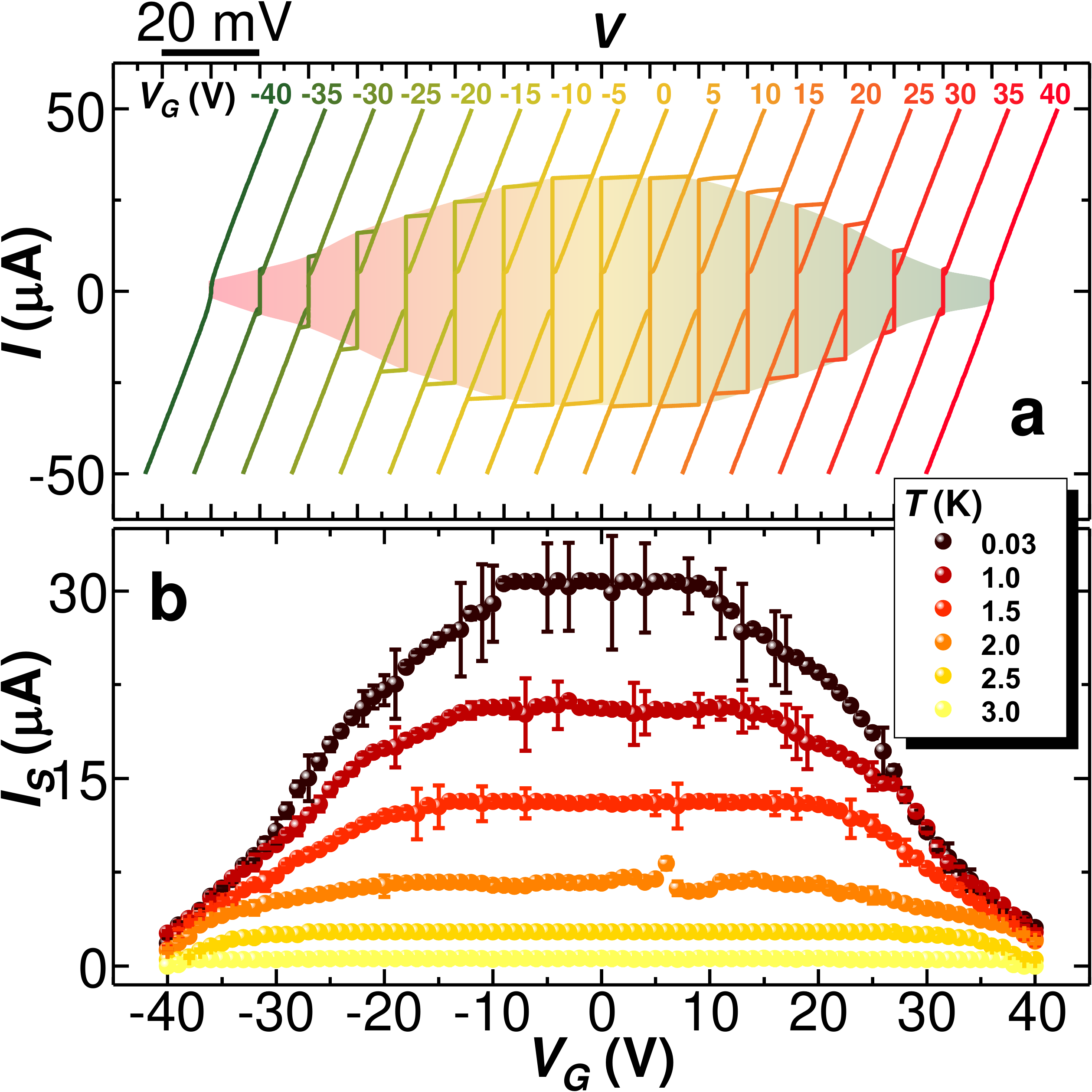}
\caption{\label{fig2} (a) Current $I$ \textit{vs.} voltage $V$ characteristics at several gate voltages $V_G$ for the same Nb-FET of Fig. \ref{fig1}. Curves are horizontally offset for clarity. A clear bi-polar suppression of the switching current is visible as $|V_G|$  is increased. (b) $I_S$ \textit{vs.}$V_G$ for  different temperatures $T$ ranging between 30 mK and 3 K. $I_S$ values were collected by measuring 50 repetitions of the $I(V)$ characteristics. Error bars represent the standard deviation of the samples.}
\end{figure}

Our Nb-FETs were fabricated by a  single-step electron-beam lithography (EBL) to pattern a polymethyl-methacrylate (PMMA)/Al (thickness: 250 nm/11 nm) bi-layer mask on a sapphire substrate. A 10-nm-thick Ti adhesion layer was then dc-sputtered, followed by 40 nm of Nb and a final metal lift-off procedure in acetone bath. Since Nb is a refractory material, its evaporative deposition is usually not performed to avoid high crucible temperatures and the following out-gassing of the organic mask,  which results in  a  reduction of the film quality \cite{Wei2008,Ohnishi2008}. Usual nano-patterning of Nb films relies on sputter deposition followed by EBL and etching, but, to avoid residues originating from the reaction between the etching gases and the PMMA, we opted for an EBL procedure followed  by sputter deposition and lift-off. 

Figure \ref{fig1}(b) and (c) show false-color scanning electron micrographs taken with different magnifications of a representative Nb-FET. The Nb DBs [colored in blue in Fig. \ref{fig1}(b) and (c)] are about  $\sim90$ nm wide, 100 nm long, and  have a normal-state resistance $R_{DB}\sim 30$ $\Omega$.  The gate [shown in green in Fig. \ref{fig2}(b) and (c)], is separated from the weak link by a $\sim70$-nm-wide gap. The results presented in the following were obtained on the same device, measured in a filtered cryogen-free dilution refrigerator at temperatures down to 30 mK. The biasing scheme of the Nb-FET is depicted in Fig. \ref{fig1}(b). 

From the measurement of the resistance ($R$) \textit{vs.} temperature $T$ [see Fig. \ref{fig1}(a)] we extracted the critical temperature of the Nb film, $T_{Nb}\simeq 7.9$ K, corresponding to a zero-temperature BCS energy gap $\Delta_0=1.764 k_B T_{Nb}\simeq 1.2$ eV, where $k_B$ is the Boltzmann constant. $T_{Nb}$ is $\sim 15\%$ lower than Nb bulk critical temperature likely due to the inverse proximity effect from the Ti sticking layer. 
Due to its lateral size, the critical temperature of the DB ($T_{DB}$) turns out to be approximately one half of that of the pristine film.  
Below $T_{DB}$, dissipationless charge transport  occurs: the  current-voltage  ($I-V$)  characteristics, recorded at temperatures ranging between 30 mK and 6.9 K are shown in Fig. \ref{fig1}(d). A switching critical current $I_S\sim 30$ $\mu$A was observed at $30$ mK, displaying the usual hysteretic behavior which stems from heating dissipated within the wire while switching from the resistive to the superconducting state \cite{Skocpol1974,Courtois2008}. 
The decay of $I_S$ \emph{vs.} $T
$ is shown in Fig. \ref{fig1}(e) along with a fit of the Bardeen equation (black dotted line), $I_S(T) =I_{0c}[1-(\frac{T}{T_{DB}})^2]^{3/2}$, where $I_{0c}=(30.0\pm 0.1)$ $\mu$A and $T_{DB}=(3.16\pm 0.01)$ K are the  zero-temperature DB critical current and temperature derived from the fitting procedure \cite{Bardeen1962}, respectively. The behaviour of the retrapping current ($I_R$) is also shown. Above the threshold temperature, $T_h\sim 2.5$ K, the hysteretic behaviour disappears, and $I_R$ coincides with $I_S$.
\begin{figure}
\includegraphics[width=0.9\columnwidth]{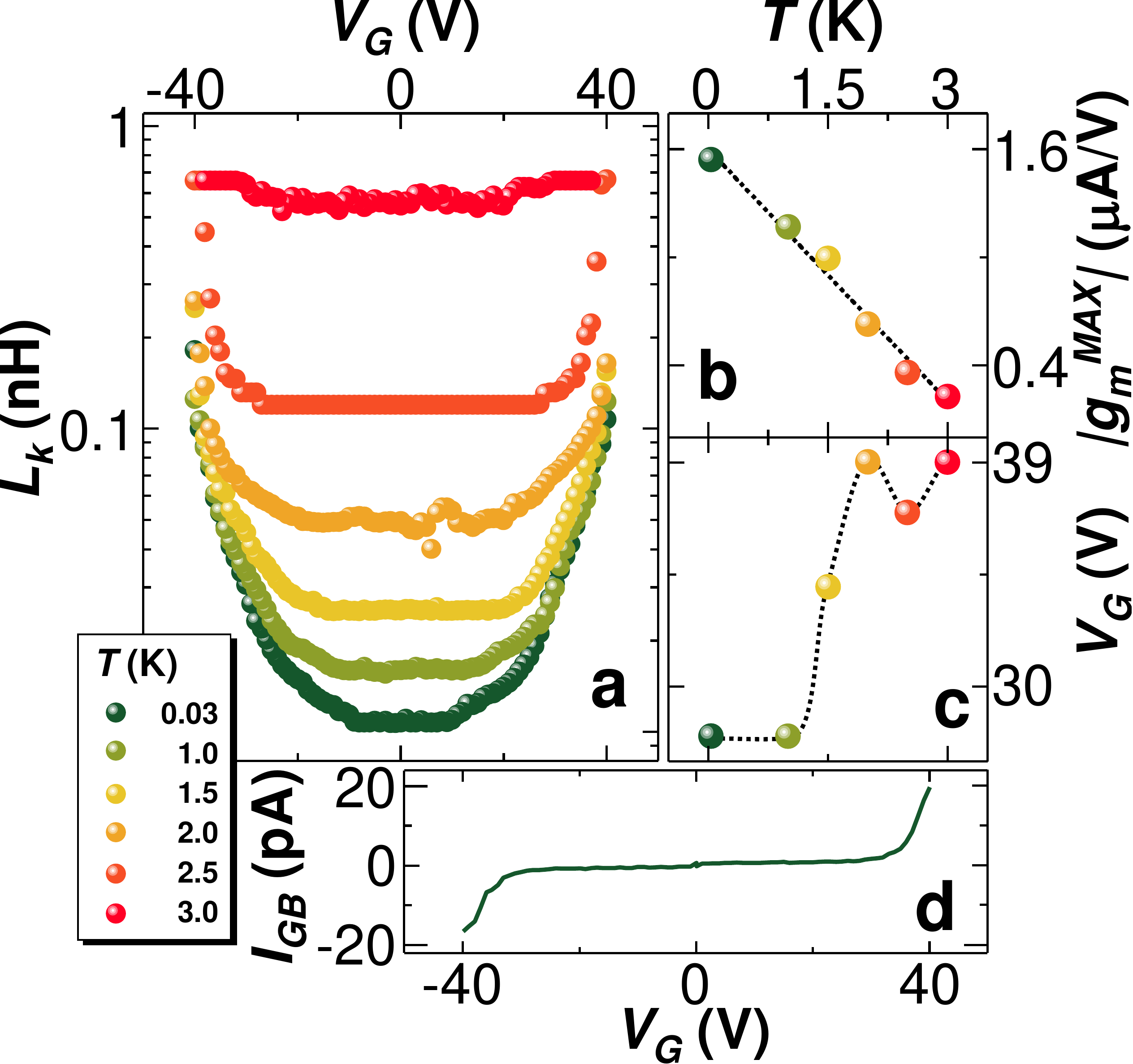}
\caption{\label{fig3} (a) Kinetic inductance $L_k$  \emph{vs.} gate voltage $V_G$ for a Nb-FET at several temperatures $T$. 
Data were deduced from the expression $L_k=\hbar/2eI_S$. 
(b) Maximum of the absolute value ($g_m^{MAX}$) of the transconductance $g_m=dI_s/dV_G$ \emph{vs.} $T$. Data were determined from the numerical derivative of the data shown in Fig. \ref{fig2}(b). 
(c) Value of gate $V_G$ at which the maximum of transconductance $\left| g_m^{MAX}\right|$ occurs as a function of  $T$. 
(d) Gate-DB current $I_{GB}$ \emph{vs.}  $V_G$ at 30 mK.}
\end{figure}

The investigation of field effect in Nb-FETs was performed by measuring $I_S$  \textit{vs} gate  voltage  $V_G$  (see Fig. \ref{fig1})\cite{DeSimoni2018,Paolucci2018,Paolucci2019,DeSimoni2019,Paolucci2019a,Puglia2019,Paolucci2019b,Bours2020}.  
Figure \ref{fig2}(a) displays the DB transistor $I-V$ characteristics measured at 30 mK for selected values of $|V_G|$ increasing up to 40 V.
The critical current $I_S$ displays a plateau at low $V_G$ values, then monotonically decays  by  increasing $|V_G|$ reaching, at 30 mK, a suppression of about $90\%$ with respect to the unperturbed value. 
Yet, as already reported on similar setups \cite{DeSimoni2018,Paolucci2018,Paolucci2019,DeSimoni2019,Paolucci2019a,Puglia2019,Paolucci2019b,Bours2020}, the electric field does not affect the transistor normal-state resistance $R_{DB}$. 
The  full  temperature  dependence  of  field  effect  is shown in Fig. \ref{fig2}(b), which displays the $I_S$ \textit{vs.} $V_G$ characteristics for selected bath temperatures up to 3 K. By increasing $T$ the $I_S$ plateau widen, but their suppression is still visible up to 3 K.  
Notably, when $T\gtrsim 2$ K, full suppression of $I_S$ was observed for $V_G > 40$ V. 
Moreover, $I_R$ is not affected by $V_G$ until it coincides with $I_S$ due to the action of either gate voltage or temperature\cite{DeSimoni2018,Paolucci2018,Paolucci2019,DeSimoni2019,Paolucci2019a,Puglia2019,Paolucci2019b,Bours2020}. 
This latter consideration is relevant in view of a possible implementation of Nb-based EF-Trons operating at 3K, where the absence of the hysteretic behavior might allow for fast gate-driven switching between the normal and the superconducting state.

We now turn to discuss some figures of merit which are relevant for possible applications of the Nb-FETs. The  kinetic  inductance $L_k=\hbar/2eI_S$ (where $\hbar$ is  the Planck constant and $e$ the unitary charge) is  the quantity usually  analyzed  in  Josephson  junctions,  and  plays a fundamental role in applications requiring non galvanic read-out of the junction state. Figure \ref{fig3}(a) shows the $L_k$ \textit{vs.} $V_G$ characteristics, calculated from the $I_S$ measurements, at several different temperatures. 
The maximum value of zero-gate kinetic inductance is $L_k\sim 0.7$ nH obtained at 3 K, while for fixed temperature, gate-dependent modulations of $L_k$ ranges from $\sim$1 nH at 30 mK up to $\sim100$ pH at 3 K. 
Such behaviour originates from the lower gate-dependent variation of $I_S$ at higher temperatures, and reflects also in the evolution in temperature of the gate-channel transconductance, which is defined for a S-FET as $g_m=dI_S/dV_G$. 
To highlight the temperature dependence of the  transconductance, the absolute value of its maximum ($|g^{MAX}_m|$) is plotted as a function of the temperature in Fig. \ref{fig3}(b).  
$|g^{MAX}_m|$ linearly decreases as a function of $T$. 
By contrast, stemming from the widening of the $I_S$ \textit{vs.} 
$V_G$ plateau, the gate voltage at which the maximum occurs $V_G^{MAX}$ increases \textit{vs.} $T$ [see Fig. \ref{fig3}(c)]. The maximum value of $|g^{MAX}_m|\simeq 1.6$ $\mu$A/V was obtained at 30 mK. Remarkably, at $3$ K, i.e., just below $T_{DB}$, $|g^{MAX}_m|$ is still equal to 0.3 $\mu$A/V. 
To provide the reader with a term of comparison, we remind that such values are a few orders of magnitude larger than those achievable in semiconductor nano-wire Josephson transistors \cite{Doh2005a}, which in turn operate below $\sim 100$ mK. 

As the last figure of merit, we discuss the gate-DB current $I_{GB}$ as a function of gate voltage $V_G$. It provides information on the quality of insulation between the gate and the weak link, and allows to exclude direct injection of hot electrons into the superconducting DB. Current injection, indeed, could result detrimental for the performance of the device, leading to a substantial reduction of input-output isolation of the FET. $I_{GB}(V_G)$ was acquired at 30 mK with a two-wire technique, by using a low-noise voltage source and $10^{-11}$-gain current pre-amplifier [see Fig. \ref{fig3}(d)]. 
$I_{GB}$ is an odd function of $V_G$ exhibiting a clear threshold ($\sim35$ V) behaviour, and reaching the maximum value of $\sim 20$ pA at  $V_G=40$ V, which corresponds to $\sim10^{-7} I_S(V_G=0)$. 
Furthermore, the gate-channel transimpedance at $V_G=27$ V, i. e., where $I_S(V_G)\sim 0.5 I_S(V_G=0)$, is approximately $\sim 24$ T$\Omega$. As discussed also elsewhere (see, e. g., the Supplementary Information of Ref. [\cite{DeSimoni2018}] and the Appendix of Ref.  [\cite{Puglia2019}]), such a behaviour is hardly compatible with a
conventional hot-electron injection into the DB and, at the same time, confirms the good electrical insulation between the transistor channel and the gate. 
In addition, we measured the variation of $I_{GB}(V_G)$ on temperature, and found basically no dependence up to  4 K. %, as summarised in Table \ref{tab:table1}.
The above result suggests that at least some fraction of the measured current might  originate from dispersion in the lines of our measurement setup, since electron injection either through vacuum or the substrate is expected to be strongly enhanced as the bath temperature is increased by two orders of magnitude.
\begin{figure*}
\includegraphics[width=0.9\textwidth]{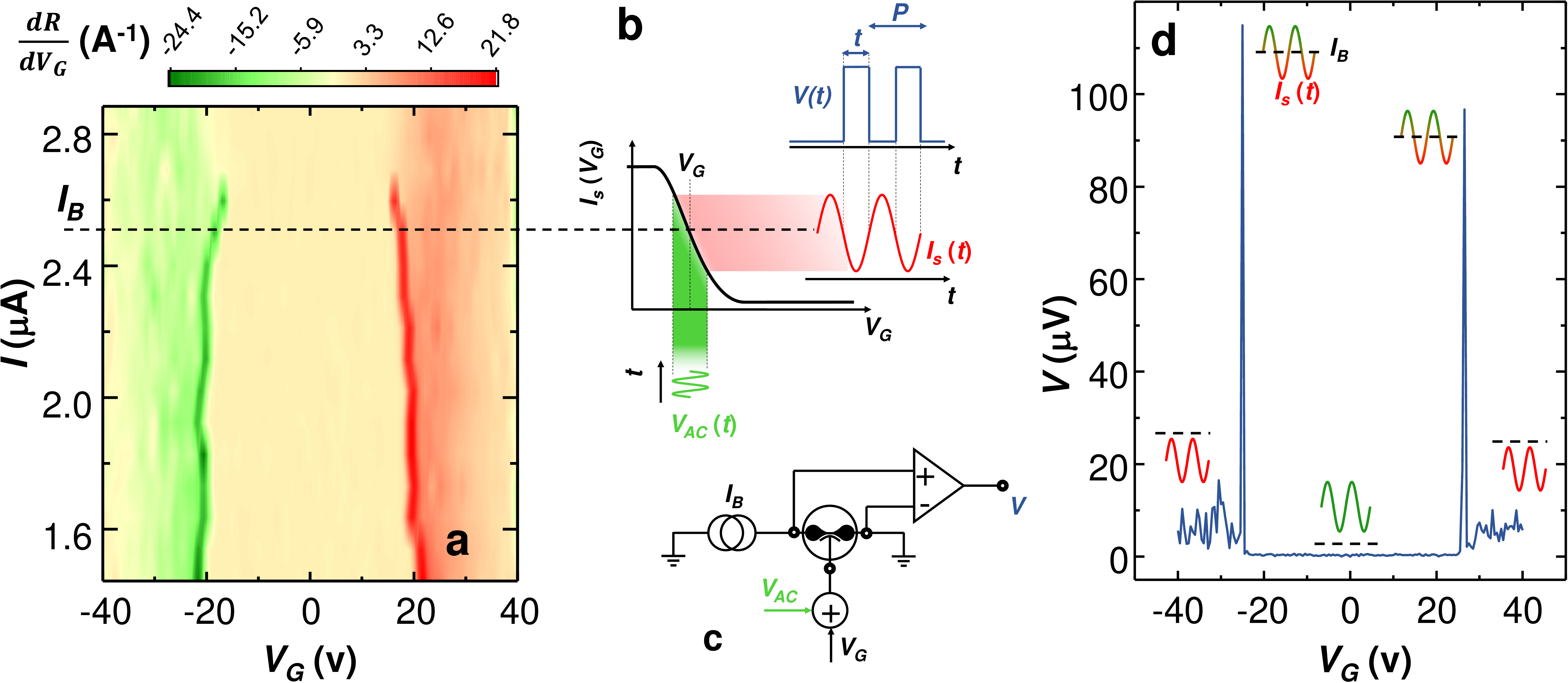}
\caption{\label{fig4} (a) Color plot of the differential resistance $dR/dV_G$  \textit{vs.}  $V_G$ and  $I$ of a Nb-FET measured at 3 K. Green and red stripes correspond to the the gate-driven superconducting-to-normal transitions of the DB. 
(b) Scheme of the operation principle  of the Nb-FET half-wave rectifier. The device is operated at constant current bias $I_B$ [black dashed line in panels (a) and (b)], whereas the gate electrode is biased with a signal composed by an ac component $V_{AC}$ (green line) and a dc component $V_G$. This results in a time-dependent switching current $I_S(t)$ (red line), which, depending on the amplitude of $V_{AC}$ and on the set point of $V_G$, yields periodic  normal-to-super and super-to-normal state transitions. In the latter condition, the voltage drop $V$ at the ends of the DB oscillates between a low and a high state (blue line) with periodicity $P$ equal to that of $V_{AC}$, and duty cycle $\tau / P$. 
(c) Biasing scheme used to implement a \emph{half-wave} rectifier based on a Nb-FET. 
(d) Voltage drop $V$ across the DB measured in a 4-wire configuration with a lock-in amplifier \emph{vs.} $V_G$.  $V_{AC}$ is the reference signal of the lock-in amplifier. $I_B$ was set to 2.5  $\mu$A.  As shown in Fig. \ref{fig4}(d), $V$ is almost zero until $I_S(V_G)<I_B$. The peaks correspond to the rectification of the ac gate signal.}
\end{figure*}

In this last section, we show the practical realization of a scheme based on a Nb-FET, which realizes a possible building block to implement a superconducting diode. 
First of all, we begin our discussion by highlighting the sharp dependence of the DB resistance $R$ on $V_G$. 
Figure \ref{fig4}(a) shows a color plot of the derivative of the four-wire transistor resistance $\frac{dR}{dV_G}$ as function of bias current $I$ and  gate voltage $V_G$ at $3$ K. 
The green and red stripes correspond to the transition to the normal state as the $I_S$ was lowered below $I_B$ due to the action of the gate voltage. 
The sharpness of the super-to-normal state transition  is a typical feature of superconducting devices that is usually and widely exploited, for instance, in transition-edge sensors (TESs) to reveal a tiny incoming radiation heating the superconductor above its $T_C$. By contrast, in our devices,  the transition events are triggered  and controlled by an electrical gate signal. 
In Fig. \ref{fig4}(b), we schematize  how to exploit gate-driven state-transitions to rectify an alternate voltage signal $V_{AC}$ applied to the gate electrode. 
$V_{AC}$ [green curve in Fig. \ref{fig4}(b)] is summed to the direct-current (dc) $V_G$ signal, %, if the temperature is opportunely set above $T_h$
and $I_S$ [red curve in Fig. \ref{fig4}(b)] is thereby modulated in time accordingly to $V_{AC}$ above and below $I_S(V_G)$. 
Therefore, depending on the constant current $I_B>0$ [see dashed black line in Fig. \ref{fig4} (a) and (b)], $I_S$ oscillates above and below $I_B$ resulting in periodic normal-to-super and super-to-normal transitions. 
The resulting voltage signal $V(t)$ across the DB [see blue curve in Fig. \ref{fig4}(b)] has the \emph{same} period $P$ of $V_{AC}(t)$, and a duty cycle $\tau /  P$ given by the time for which $I_S<I_B$. $V(t)$ oscillates between a \textit{low-state}, where $V(t)=0$ (superconducting state), and a \textit{high-state}, where $V(t)=R\cdot  I_B>0$. Such a circuit, sketched in Fig. \ref{fig4}(c), realizes a \emph{half-wave} rectifier which could be exploited in a superconducting diode, for instance, to rectify the radiation picked-up by an antenna coupled to the gate electrode. In the latter case, the sensitivity depends on the width of the switching current probability distribution of the DB [see Ref. [\cite{Puglia2019}]), while the amplitude of $V(t)$ can be enhanced by increasing $I_B$. 
With respect to the cut-off frequency of the Nb-FET-based rectifier, we note that the upper limiting frequency set by $f_\Delta$ might be reduced by the typical time-scale of the electrically-driven phase transition in S-FETs which is currently  totally unknown, and demands for a future investigation. Yet, the DB-gate capacitance is low enough ($\sim 0.1$ fF) not to play any role. 

To provide a preliminary demonstration of the rectifying behavior of the DB, we biased the gate of our Nb-FETs  according to the scheme of Fig. \ref{fig4}(c). $V_{AC}$ was provided by the sinusoidal reference of a lock-in amplifier (the frequency and amplitude of $V_{AC}$ were $\sim17$ Hz and 10 mV, respectively), while $I_B$ was kept equal to 2.5  $\mu$A. 
The voltage $V$ was measured in-phase as a function of $V_G$. As shown in Fig. \ref{fig4}(d), $V$ is almost 0 until $I_S(V_G)<I_B$. When $I_S(V_G)$ crosses $I_B$, at $V_G\sim24$ V, sharp peaks appear, corresponding to the differential resistance peaks, and demonstrate the occurrence of a rectified in-phase voltage signal across the DB. By further increasing $V_G$, $V$ drops to lower values since  $I_S(V_G)$ is constantly lower than $I_B$. 
In this $V_G$ configuration, nonetheless, $V$ is never equal to zero  due to the gate-dependent DB resistance \cite{Paolucci2018}.

In summary, we have demonstrated Nb-based \emph{all-metallic} Josephson field effect transistors operating up to $\sim 3$ K, which could be pivotal for the implementation of superconducting digital logic ports. Our nano-bridge showed full quench of the Josephson current due to the application of a gate-voltage $V_G$. 
The dependence of the kinetic inductance and of the   transconductance on $V_G$ suggest that these nano-devices are competitive with respect to conventional semiconductor nanowire-based Josephson transistors. 
We have finally also demonstrated the operation of a superconducting half-wave rectifier to be exploited either in SCE or for photon detection applications.

The  authors  acknowledge  the  European Union’s Horizon 2020 research and innovation programme under grant No. 777222 ATTRACT (ProjectT-CONVERSE), and under grant No. 800923-SUPERTED.

%\appendix

%\section{Appendixes}

\nocite{*}
\providecommand{\noopsort}[1]{}\providecommand{\singleletter}[1]{#1}%

\end{document}